# Comparing the Mechanical and Thermodynamic Definitions of Pressure in Ice Nucleation

P. Montero de Hijes[1,*], K. Shi[2,*,†], C. Vega[3], and C. Dellago[1,†]
[1] *Faculty of Physics, University of Vienna, A-1090 Vienna, Austria*
[2] *Department of Chemical and Biological Engineering, University at Buffalo,
The State University of New York, Buffalo, NY 14260, USA* and
[3] *Departamento de Química Física, Facultad de Ciencias Químicas,
Universidad Complutense de Madrid, 28040 Madrid, Spain*

Crystal nucleation studies using hard-sphere and Lennard-Jones models have shown that the pressure within the nucleus is lower than that in the surrounding liquid. Here, we use the mechanical route to obtain it for an ice nucleus in supercooled water (TIP4P/Ice) at 1 bar and 247 K. From this (mechanical) pressure, we obtain the interfacial stress using a thermodynamic definition consistent with mechanical arguments. This pressure is compared with that of bulk ice at equal chemical potential (thermodynamic pressure), and the interfacial stress with the interfacial free energy. Furthermore, we investigate these properties on the basal plane. We find that, unlike in hard-sphere and Lennard-Jones systems, mechanical and thermodynamic pressures agree for the nucleus, and the interfacial stress and free energy are comparable. Yet the basal interface displays an interfacial stress nearly twice its interfacial free energy, suggesting that this agreement may be system-dependent, underscoring the limitations of mechanical routes to solid–liquid interfacial free energies.

*These authors contributed equally to this work
†Corresponding author:
kaihangs@buffalo.edu
christoph.dellago@univie.ac.at

Ice nucleation is a critical process in nature and many technological fields [1, 2]. However, investigating critical nuclei is difficult as nucleation events may take seconds to occur but may last only for a few nanoseconds. Computer simulations have significantly contributed to our understanding of this process [3]. In particular, molecular dynamics simulations in the NVT ensemble provided a means to, not only, simulate the process but also to stabilize critical nuclei [4, 5], allowing us to study them in great detail. This has made it possible to carefully study the critical crystal nuclei of hard spheres (HS)[6, 7] and Lennard-Jones (LJ) particles[8], showing that the pressure inside is smaller than outside. Indeed, earlier work on hard spheres with short-range attractive interactions [9] and on binary mixtures [10] had implicitly indicated this, as reflected in the reported densities. These observations are, a priori, in contradiction with the Young-Laplace equation,

$$\Delta p^\mu = \frac{2\gamma}{R}, \quad (1)$$

where $\Delta p^\mu$ is the difference in pressures assuming bulk phases ($\Delta p^\mu = p^\mu_{\text{I}_\text{h}} - p_\text{w}$ for ice Ih and water, respectively, in this work) and $\gamma$ is the interfacial free energy for a dividing surface defined as the surface of tension located at $R$. The apparent conflict resolves once we recognize that the thermodynamic behavior of crystal nuclei differs from that of the bulk crystal. Since Eq. 1 employs reference bulk states, it does not matter that at a given chemical potential and temperature, the pressure of the nucleus differs from that of the bulk [6]. The true pressure of the nucleus is not directly related to $\gamma$ but, instead, it is related to the interfacial stress $f$ [7]. Similar to the Young-Laplace equation, $f$ can be defined thermodynamically as [7, 11, 12]

$$\Delta p = \frac{2f}{R}, \quad (2)$$

where $\Delta p = p_{\text{I}_\text{h}} - p_\text{w}$ is the actual difference in pressure between the two phases. The origin of discrepancies in pressure between an ideal bulk crystal and the core of the nucleus at the same $\mu$ is likely on the concentration of defects and possibly the presence of strain [7, 12, 13].

The radii of curvature $R$ in both Eqs. 1 and 2 are often assumed to be equal for simplicity but this is an approximation. Moreover, a mechanical route to $\gamma$ in solid-liquid interfaces has not been successful to date [3] while several expressions have been proposed for $f$ [7, 8, 14, 15]. At the root of all this uncertainty is the arbitrariness in both i) the location of the dividing surface in Gibbs' thermodynamics and ii) the definition of the pressure tensor in mechanics [16]. Furthermore, the fact that in fluid-fluid interfaces $\gamma = f$ (except for small droplets or confined systems [17–25]) has often led to misunderstandings of the solid-liquid interface. Nevertheless, Gibbs already noted that the tension of the surface —meaning $\gamma$— did not refer to the true tension —meaning $f$—. However, Gibbs wrongly believed that the differences would be negligible in most cases [26].

Despite some remaining uncertainties, our understanding of solid-liquid interfaces has significantly grown thanks to computer simulations, which have allowed researchers to confirm that $f$ is often negative in solid-liquid planar [27–32] and spherical [3, 6, 8] interfaces, drastically differing from $\gamma$, which is always a positive property. Moreover, systems may present $f < 0$ in some crystallographic planes while in others $f > 0$ [33, 34]. Mechanically, $f$ is rooted in the stress profile $S(r) = P_N(r) - P_T(r)$, where $P_N(r)$ and $P_T(r)$ are, respectively, the normal and tangential components of the pressure tensor and $r$ is the axis normal to the interface. Notably, $S(r)$ can be nonuniform with both positive and negative contributions depending on the interfacial layer [35, 36]. This nonuniformity was observed also in the water/vapor interface [37], even though in fluids $f$ typically coincides with $\gamma$. Becker et al. [38] suggested that the density difference between the two phases and the bonding energy play an important role in $f$, whereas Eriksson and Rusanov [39] hypothesized that solid-liquid interfaces with high mobility could lead to $f = \gamma$. Recent work indicates that, in a few cases, this could be the case for the planar ice-water interface. In Ref. [40] a thermodynamic formalism for fluid interfaces was reasonably successful in describing the planar ice-water interface using TIP4P/Ice [41]. In contrast, studies using the mW model [42] revealed in Ref. [32] that while $f$ and $\gamma$ are the same at one particular point of the coexistence line, they differ for most melting temperatures.

Here, we test whether the significant discrepancy between $f$ and $\gamma$ during nucleation observed in simple models also occurs for an important substance like water. In addition, we compare these two quantities for the basal plane at coexistence. We use the TIP4P/Ice water model [41] simulated using GROMACS-2021.3-mixed with (i) a time step of 2 fs; (ii) the Nosé-Hoover thermostat (relaxation time of 1 ps) fixed at 247 K (nucleus) and 270 K (basal plane); (iii) the Parrinello-Rahman barostat (relaxation time of 2 ps) fixed at 1 bar (isotropic for the nucleus and $Np_yT$ for the basal plane); (iv) the particle-mesh-Ewald algorithm of order 4 and Fourier spacing of 0.1 nm; (v) a cutoff of 0.9 nm in both Lennard-Jones and Coulomb interactions; (vi) long-range corrections to the Lennard-Jones term. The dispersion correction to the pressure yields $-350.9$ bar in the nucleus and $-337.4$ bar in the basal plane.

Following previous works on hard spheres and Lennard-Jones, we attempted to stabilize an ice nucleus in the NVT ensemble [5–8]. However, after some failed attempts, we opted for generating trajectories in the NpT ensemble from a critical nucleus determined previously via the seeding approach [43], and looking for paths where the nucleus is stable long enough to gather data. We found 6 trajectories out of 40 where the nucleus survived for about 50 ns each at 247 K and 1 bar, allowing us to gather 300 ns. The system is shown in Fig. 1 and contains 78,856 molecules, of which ∼1,500 belong to the nucleus. For the planar interface, the system contained 20,736 molecules, about half being ice and half being liquid. The cross sectional area defining the interface is determined to provide bulk behavior in the ice far from the interface.

The interfacial free energy $\gamma$ has been widely investigated for TIP4P/Ice water model [40, 44–46]. Therefore, we estimate $\gamma$ from the supercooling $\Delta T$ using the empirical fit suggested in Ref. [40]. Then, using Eq. 1, we obtain the pressure $p_{I_h}^\mu$ of bulk ice at the same chemical potential as the external liquid, which is homogeneous throughout the whole system. To do so, we first need to compute both the water pressure $p_w$, which is given by the average pressure in the system (see supplementary material in Ref. [6]), and the radius of the nucleus $R$. In the introduction, curvature effects in $\gamma$ due to the location of the dividing surface were not described for simplicity. Nevertheless, to be precise, we should denote $\gamma$ and $R$ in Eq. 1 as $\gamma_s$ and $R_s$ indicating that they are defined at a particular location of the dividing surface within the interfacial region, i.e. the surface of tension [47, 48], at which the derivative of $\gamma$ with respect to the position of the dividing surface vanishes (see Supporting Information for further details). Determining $R_s$ requires free energy calculations to find the minimum in $\gamma[R]$, where the square brackets imply a variation with the location of the dividing surface, not with the actual radius. However, it has been shown empirically that a particular criterion based on the order parameter $\bar{q}_6$ [49] to classify ice-like and water-like molecules, allows us to obtain a radius leading to agreement in free energy barriers with those obtained from free energy calculations [43]. We refer to Refs. [40, 45, 50] for details on this criterion. The number of ice-like molecules $N_{I_h}$ is estimated from the $\bar{q}_6$ using a threshold of 0.365 and a cutoff of 3.5 Å. The cluster radius $R_s$ is then obtained from:

$$R_s = \left(\frac{3N_{I_h}}{4\pi\rho_{I_h}^\mu}\right)^{(1/3)}, \quad (3)$$

where $\rho_{I_h}^\mu$ is the number density of bulk ice (i.e. the number of water molecules per unit of volume of ice Ih for TIP4P/Ice at 247 K and $p_{I_h}^\mu$). Since the density of ice barely changes with pressure (∼0.5% over 500 bars [45, 51]) we take it from bulk ice at 247 K and 1 bar even though bulk ice at such conditions will not have the same $\mu$ as the liquid but lower. The error introduced by this approximation is certainly smaller than the one introduced in $R_s$ via an empirical definition.

In contrast, $p_{I_h}$ can be measured directly from molecular dynamics trajectories using the virial (or mechanical) route [16]. Although the pressure tensor at a point is not uniquely defined, Shi et al. [52] recently showed that it is possible to define a unique pressure tensor over a small region of space, roughly the range of the intermolecular forces, in a planar geometry. However, the validity of such a unique definition remains unclear in systems with curved interfaces. Here, we are interested in the local pressure of the nucleus far enough from the interface where the ambiguity in the pressure definition is of negligible order. In this work, we adopted the contour definition of Irving and Kirkwood [14, 53], which is a straight line connecting two interacting molecules. All parameters in the pressure tensor calculations are consistent with those in the molecular dynamics simulations, except for the Coulombic interactions. Incorporating the Ewald summation into the pressure tensor formulation is not trivial [54]. Instead of implementing the Ewald summation directly, we adopted the shifted-force version [55, 56] of the Wolf method [57] to account for the long-range contribution to the local pressure tensor. This allows us to calculate the pressure tensor in a pairwise manner with accuracy comparable to the exact Ewald method. Derivation of molecular pressure tensor equations in spherical coordinates, calculation details, and source code are available in the Supporting Information.

From the pressure profiles, we measure the internal pressure $p_{I_h}$, and then $f$ is estimated from Eq. 2 upon defining $R$. Just like with $\gamma$, the arbitrariness in the location of the dividing surface is expected to affect $f$. In the Supporting Information we show that $R$ in Eqs. 1 and 2 is not necessarily equal. The former, $R = R_s$, is the surface of tension defined by Gibbs, whereas the latter, $R = R^*$, is the true surface of tension. There are, however, mechanical definitions of the surface of tension that depend on the contour definition of the local pressure tensor [11, 14]. The latter is non-unique, thus the mechanical definition of the true surface of tension may be non-unique. Nevertheless, a mechanical definition of $f$ consistent with Eq. 2 that is invariant to the local pressure tensor choice and allows us to define a unique true surface of tension is given by

$$f = R^* \int_0^\infty dr \frac{1}{r}[P_N(r) - P_T(r)], \qquad (4)$$

(see Ref. [14] and Supporting Information for further details). How to evaluate $R^*$ is, however, non-trivial and no empirical rules have been suggested. Since the uncertainty in $R^*$ must be comparable to the interfacial thickness, we pragmatically estimate $f$ for three different values including $R_s$ and the two limits of the interfacial region.

We proceed now first by describing the nucleus via the Young-Laplace equation (Eq. 1). Using the seeding technique, we identified that it was a critical nucleus at 247 K and 1 bar. Since the melting point $T_m$ of the model is 270 K, we use the empirical fit from Ref. [40], $\gamma = 26.6 - 0.174 \cdot \Delta T$ (mJ/m$^2$), where $\Delta T = T_m - T$, to obtain $\gamma \sim 23$ mJ/m$^2$. The pressure of the liquid equals that of the barostat, hence, $p_w = 1$ bar. For obtaining the number density, we assume the incompressibility approximation is valid and we use that of bulk ice at 247 K and 1 bar after checking that a 500 bar difference leads to a negligible density change (within 0.5%)[45, 51]. The mass density of bulk ice at 247 K and 1 bar is $\sim 0.91$ g/cm$^3$. Hence, the number density $\rho_{I_h}^\mu \sim 30.5$ nm$^{-3}$. As shown in Fig. 2, $N_{I_h} \sim 1500$ molecules and, finally, $R_s$ is estimated from Eq. 3 to be $\sim 2.3$ nm. From these estimates and using Eq. 1, we find that $p_{I_h}^\mu \sim 200$ bar. This way of defining the nucleus pressure was referred to as thermodynamic pressure in the context of hard spheres in Ref. [6]. Note that the thermodynamic pressure is always larger than the external one. How far is this representation from the true (also called mechanical) pressure of the nucleus? How does $\gamma$ compare to $f$? Gibbs believed that they would not be too different. However, in simple systems like hard spheres [6, 7] and Lennard-Jones [8], this is far from true since it has been observed that $p_{I_h} < p_w$, leading to $f < 0$, even though $\gamma > 0$ by thermodynamic definition.

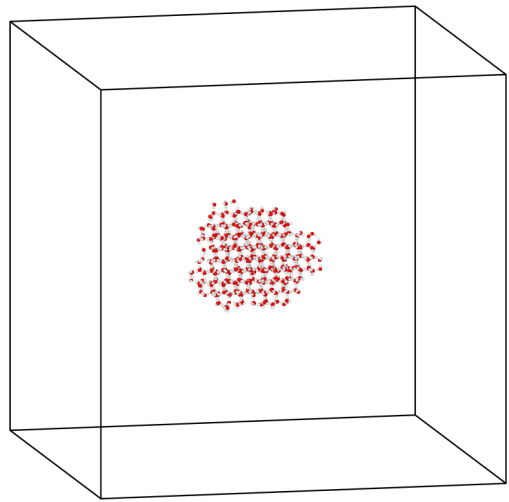

FIG. 1. Snapshot of the critical nucleus configuration showing only the molecules belonging to the nucleus.

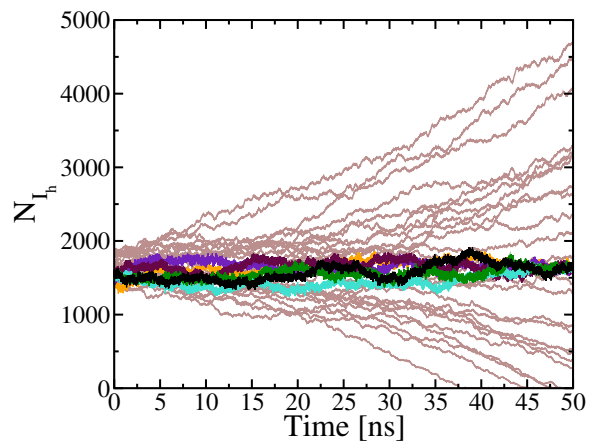

FIG. 2. Time evolution of the number of ice-like molecules, $N_{I_h}$ for the system in Fig. 1. Brown lines show trajectories where the nucleus transited too early to acquire data, whereas the other colors show the trajectories that were finally employed for our analysis.

To answer these questions for the ice-water interface, we now analyze the trajectories to collect microscopic information. First, the nucleus' center of mass (COM) is determined at every step, and from the nucleus' COM toward the boundary of the simulation box, we measure the density profile, shown in Fig. 3 a). As can be seen, the interface is not sharp. Only ice molecules within 2 nm from the COM are in a density plateau, whereas the interface spans almost the same length. Therefore, the actual radius of the nucleus is between $\approx 2$ and 3.5 nm, suggesting that our previous estimate of $R_s \sim 2.3$ nm is reasonable. Also, the actual density is in agreement with the mass density of bulk ice, 0.91 g/cm$^3$, within the uncertainty. Nevertheless, a very similar density between the nucleus and the bulk does not guarantee that the pressures will also be similar, as observed in hard spheres[3, 6]. Therefore, the true pressure of the nucleus should be computed to clarify this.

We analyze the local pressure tensor profile via the mechanical route. In this case, the tensor has two non-zero components: the normal (radial) pressure $P_N$ and the tangential pressure $P_T$ (averaged from the polar and azimuthal pressures, see Supporting Information). The tangential component is rather noisy as can be seen in Fig. 3 b) so we estimated it, $P_T$calc, also from the fit to the normal component, $P_N$fit, combined with the hydrostatic equilibrium condition as explained in the Supporting Information. As shown in Fig. 3 b), the pressure profile converges to the average pressure when it reaches the liquid far from the interface (within a 10 bar error). More importantly, it shows that the true pressure in the nucleus $p_{I_h}$ agrees with that of bulk ice at the same chemical potential as the system $p_{I_h}^\mu$, i.e. the true (mechanical) pressure inside the nucleus is $\sim$ 200 bar. Interestingly, the interface is stretched even to negative pressures. However, we cannot quantify this effect with certainty due to the non-uniqueness in the definition of the local pressure tensor. In any case, the pressure profile is expected to cover significantly different pressures from about 200 bar at the core of the nucleus, down to negative pressure at the interface, and up again to standard pressure at the surrounding liquid.

Considering the lowest (2 nm) and highest (3.5 nm) bounds for $R^*$, we find that $f$ should be between 20 and 35 mJ/m$^2$. In particular, for $R^* \approx R_s$, $f \sim \gamma \sim 23$ mJ/m$^2$. Therefore, in the critical nucleus at standard pressure and 23 K of supercooling, $f$ is comparable with $\gamma$ for the TIP4P/Ice model. According to Eriksson and Rusanov [39] the high mobility of molecules at the ice-water interface would promote the equivalence, also supported by the lower density in the internal phase according to Ref. [38]. Hence, Gibbs assertion of their equivalence is not completely invalidated; rather, the extent to which they coincide seems to be system dependent.

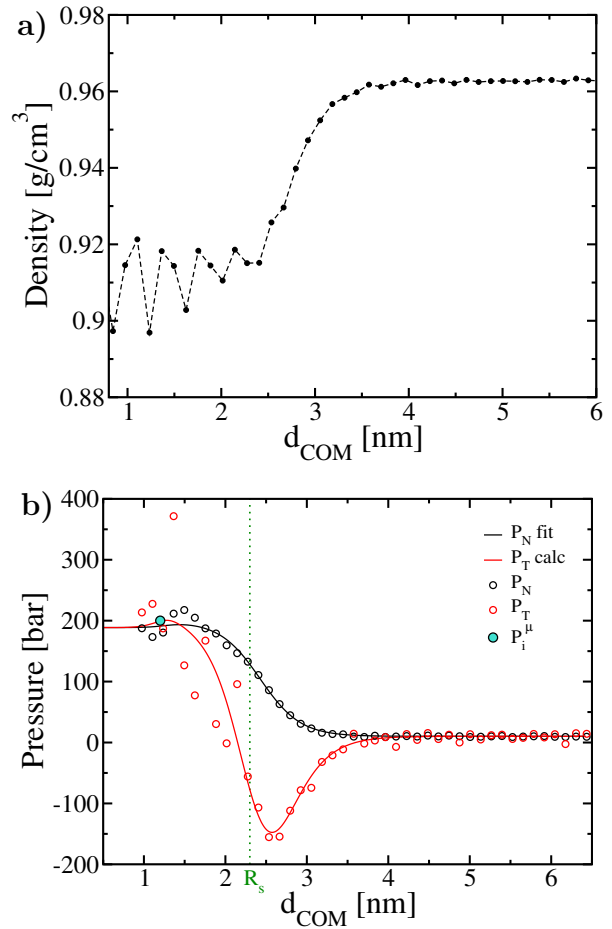

FIG. 3. a) Density and b) pressure profiles as a function of the distance $d_{COM}$ from the center of mass of the ice nucleus. In panel b), the thermodynamic pressure (blue symbol) and the position $R_s$ of the surface of tension (vertical dotted green line) are also shown.

Since the equivalence between $f$ and $\gamma$ seems to be system-dependent, we now move from the nucleus to study a just slightly different case, the planar interface at coexistence for the basal plane (Fig. 4 a), whose $\gamma_|$ is well known to be $\sim 27.2$ mJ/m$^2$ [40, 58]. In this case, the temperature is 270 K instead of 247 K, and a single plane, instead of an average of planes is exposed. As can be seen in Fig. 4 b), the density profile exhibits an interface of similar thickness to the nucleus, $\sim 2$ nm. The density profile is smoothed using a one-dimensional Gaussian filter with standard deviation $\sigma = 3$Å [59]. Moreover, in Fig. 4 c), we show the pressure profile across the basal plane for both the normal and tangential components. Normal and tangential pressures, in this case, are perpen-dicular and parallel to the planar interface, respectively. They are calculated using the pressure tensor equations from Ref. [60] (see Supporting Information). The normal pressure is constant at $\sim 7$ bar within statistical uncertainties (standard deviation $\sim 7.2$ bar), confirming that the system has reached mechanical equilibrium. Only the diagonal elements in the pressure tensor are non-zero; all off-diagonal pressure components are confirmed to fluctuate around 0. The original tangential pressure data in the ice phase are noisy due to the large fluctuation in density. To smooth the data, we apply the same one-dimensional Gaussian filter as in the density case. We also fit the raw data with a skewed Gaussian function (see Supporting Information). In this planar interface, we distinguish the interfacial stress from the spherical nucleus one with the symbol $f_|$, defined as [61]

$$f_| = \frac{1}{2} \int_0^{L_y} dy [P_N(y) - P_T(y)], \quad (5)$$

where $L_y$ is the length of the simulation box in the direction perpendicular to the interface and the division by 2 accounts for the two interfaces present in the system due to periodic boundary conditions. After applying Eq. 5 to our trajectory, we obtain $f_| \sim 50$ mJ/m$^2$, almost twice the value of $\gamma_|$. This shows that the interfacial stress and the interfacial free energy may differ in the ice-water interface. The interfacial stress seems notably more sensitive to thermodynamic changes than the interfacial free energy as previously observed using the mW model in Ref. [32]. Last but not least, note that in both the nucleus and the basal plane the interfacial stress is positive as the interface is stretched to negative pressures.

In summary, we have investigated the pressure inside a critical ice nucleus in supercooled water at 1 bar and 23 K of supercooling simulated via the TIP4P/Ice model. The planar interface for the basal plane is also studied for comparison. The mechanical pressure inside the ice nucleus is compared to the bulk ice pressure. Then the interfacial stress and free energies are also compared. Our findings contrast notably with previous studies on hard-sphere and Lennard-Jones systems. While these simpler systems showed discrepancies between mechanical and thermodynamic approaches, our results showed agreement for ice nucleation under the studied conditions.

However, we argue that it may be just coincidental, as the interfacial stress was observed to be very sensitive to changes in conditions, as evidenced by our examination of the basal planar interface at a temperature 23 K higher. There, the interfacial stress becomes approximately twice the magnitude of the interfacial free energy. Further work is needed to thoroughly elucidate the relation between $f$ and $\gamma$ during nucleation. What we have described here at 247 K and 1 bar may differ from what occurs at other temperatures and pressures.

The marked sensitivity to temperature changes of $f$

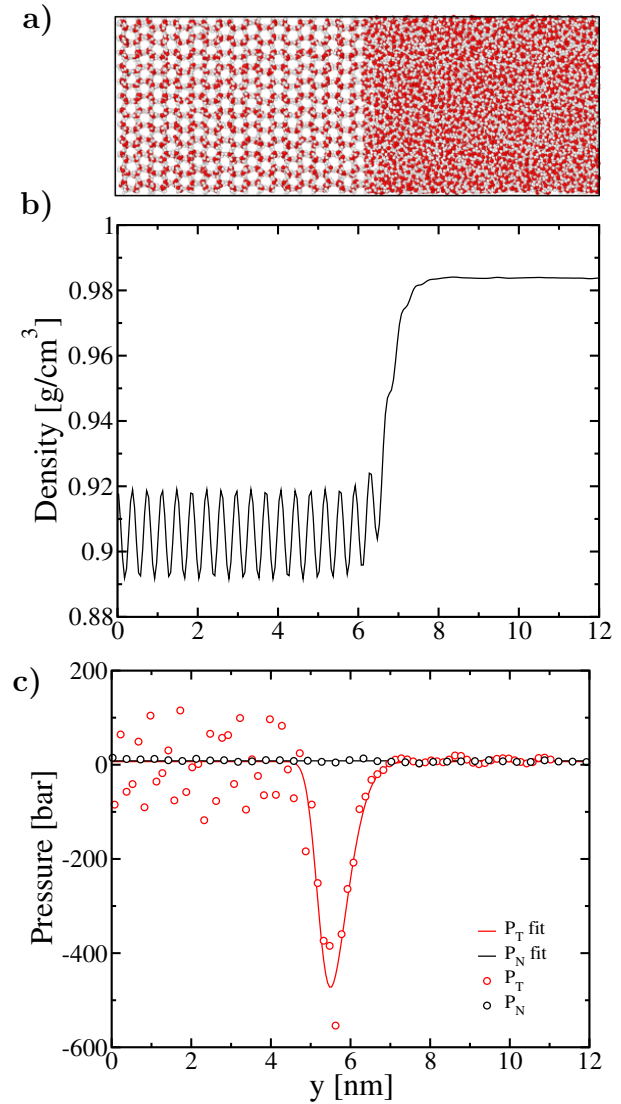

FIG. 4. a) Initial configuration of the system exposing the basal plane to the liquid and the secondary prismatic face to the reader. b) Density and c) pressure profiles across the ice–water interface exposing the basal plane at equilibrium at 270 K and 1 bar, obtained in the $Np_yT$ ensemble. Both the density and tangential pressure points shown in the figure were ones after smoothing using a one-dimensional Gaussian filter with $\sigma = 3$ Å [59]. The tangential pressure fit (line) was generated by fitting the original data to a skewed Gaussian function. Panels b) and c) share the same horizontal axis, corresponding to the position along the direction normal to the interface (y-axis in the simulation box and analysis code). Raw data for density and pressure profiles are provided in the SI.

strongly suggests that mechanical routes should not be relied upon for calculating $\gamma$, as they may lead to significant inaccuracies even under modest variations in conditions. Indeed, it is a conceptual error as the interfacial free energy is defined from bulk reference states. In most cases, the liquid can be assumed to be bulk and we have to find the reference bulk solid at the same chemical po-tential as the liquid. In confined systems [17–25], finding the reference bulk liquid may also be required.

We show that the pressure across an ice nucleus in supercooled water at 247 K may vary from an increased pressure of 200 bar in the core of the nucleus down to negative pressure at the interface before reaching the external pressure of 1 bar. Our work provides insights into the relationship between mechanical and thermodynamic properties during ice nucleation, while also highlighting the limitations of mechanical approaches for interfacial free energy calculations. We hope to motivate further work along the line of harmonizing the thermodynamic and mechanical picture of solid-liquid interfaces.

This research was funded in part by the Austrian Science Fund (FWF) through the SFB TACO 10.55776/F81. K.S. gratefully acknowledges the startup funds from the University at Buffalo. For open access purposes, the author has applied a CC BY public copyright license to any author-accepted manuscript version arising from this submission. Computer resources and technical assistance were provided by the Vienna Scientific Cluster (VSC), North Carolina State University High-Performance Computing Services Core Facility (RRID:SCR_022168), and the Center for Computational Research at the University at Buffalo. Atomic visualizations were made with OVITO [62]. The authors have no conflicts to disclose. The data to support these findings is available upon request.